\documentclass[%
groupedaddress,
preprint,
 amsmath,amssymb,
 aps,
prb,
]{revtex4-1}

\usepackage{graphicx}
\usepackage{dcolumn}
\usepackage{bm}

\begin{document}

\newcommand{\be}{\begin{equation}}
\newcommand{\ee}[1]{\label{#1}\end{equation}}
\newcommand{\bem}{\begin{eqnarray}}
\newcommand{\eem}[1]{\label{#1}\end{eqnarray}}
\newcommand{\eq}[1]{Eq.~(\ref{#1})}
\newcommand{\Eq}[1]{Equation~(\ref{#1})}
\newcommand{\vp}[2]{[\mathbf{#1} \times \mathbf{#2}]}


\title{ Superfluid spin transport  in  magnetically ordered solids}

\author{E.  B. Sonin}
 \affiliation{Racah Institute of Physics, Hebrew University of
Jerusalem, Givat Ram, Jerusalem 91904, Israel}

\date{\today}

\begin{abstract}
The paper reviews the theory of the long-distance spin superfluid transport in solid ferro- and antiferromagnets based on the analysis of the topology, the Landau criterion, and phase slips. Experiments reporting evidence of the existence of spin superfluidity are also overviewed.
\end{abstract}

\maketitle


\section{Introduction} \label{Intr}

 The phenomenon of spin superfluidity\cite{ES-78b,ES-82,Bun,Adv,BunV,Tserk,Halp,Mac,Pokr,Son17,Hoefer,BratAF} is based on the analogy of special cases of the Landau--Lifshitz--Gilbert (LLG) theory in magnetism and superfluid hydrodynamics. This analogy was clearly formulated long ago by \citet{HalHoh} in their hydrodynamic theory of spin waves.
 While in a superfluid mass (charge in superconductors) can be transported by a current proportional to the gradient of the phase of the macroscopic wave function, in a magnetically ordered medium there are spin currents, which are proportional to the gradient of the spin phase. The latter is defined as the angle of rotation around some axis in the spin space. Strictly speaking this analogy is complete only if this axis is a symmetry axis in the spin space. Then according to Noether's theorem the spin component along this axis is conserved. But possible violation of the spin conservation law  usually is rather weak because it is related with relativistically small (inversely proportional to the speed of light) processes of spin-orbit interaction. In fact, the LLG theory itself is based on the assumption of weak spin-orbit interaction \cite{LLstPh2}.

 The analogy of the LLG theory with the theory of superfluidity suggests a new useful language for description of phenomena in magnetism, but not a new phenomenon. During the whole period of spin superfluidity investigations and up to now there have been disputes about definition what is spin superfluidity. There is a school of thinking that the existence of any spin current proportional to the spin phase (rotation angle) means spin superfluidity \cite{BunV,spinY}.  This definition transforms spin superfluidity into a trivial ubiquitous  phenomenon existing in any magnetically ordered medium. A spin current proportional to the spin phase emerges in any domain wall and in any spin wave. Under this broad definition spin superfluidity was already experimentally detected beyond reasonable doubt in old experiments of the middle of the 20th century detecting domain walls and spin waves.
We use the term superfluidity in its original meaning known from the times of Kamerlingh Onnes and Kapitza: transport of some physical quantity (mass, charge, or spin) over macroscopic distances without essential dissipation. This requires a constant or slowly varying phase gradient at macroscopic scale with the total phase variation along the macroscopic sample equal to $2\pi$ multiplied
by a very large number. In examples of domain walls and spin waves this definitely does not take place. Gradients oscillate in space or time, or in both. The total phase variation is on the order of $\pi$ or much less. Currents transport spin on distances not more than the domain wall thickness, or the spin wavelength. Although such currents are also sometimes called supercurrents, we use the term supercurrent only in the case of macroscopic supercurrent persistent at large spatial and temporal scales.

The possibility of supercurrents is conditioned by the special  topology of the magnetic order parameter space (vacuum manifold).  Namely, this space must have topology of circumference on the plane. In magnetically ordered systems this requires the presence of easy-plane uniaxial anisotropy.  It is possible also in non-equilibrium coherent precession states, when spin pumping supports spin precession with fixed spin component along the magnetic field (the axis $z$). Such non-equilibrium coherent precession states, which are called nowadays magnon BEC, were experimentally investigated in the $B$ phase of superfluid $^3$He and in YIG films. \cite{BunV,Dem6} 

Spin superfluid transport (in our definition of this phenomenon) is possible as long as the spin phase gradient does not exceeds the critical value determined by the Landau criterion.   The Landau criterion checks stability of supercurrent states with respect to elementary excitations  of all collective modes. The Landau criterion determines a  threshold for the current state instability, but it tells nothing about how the instability develops. The decay of the supercurrent is possible only via phase slips. In a phase slip event a vortex crosses current streamlines decreasing the phase difference along streamlines. Below the critical value of supercurrent phase slips are suppressed by energetic barriers. The critical value of the supercurrent at which barriers vanish is of the same order as that estimated from the Landau criterion. This leads to a conclusion that the instability predicted by the Landau criterion is a precursor of the avalanche of phase slips not suppressed by any activation barrier.

The present paper  reviews the three essentials of the spin superfluidity concept: topology, Landau criterion, and phase slips. The paper focuses on the qualitative analysis avoiding details of calculations, which can be found in original papers. After the theoretical analysis the experiments supporting the existence of spin superfluidity are discussed. 

\section{Concept of superfluidity}

\begin{figure}[b]
\includegraphics[width=.95\textwidth]{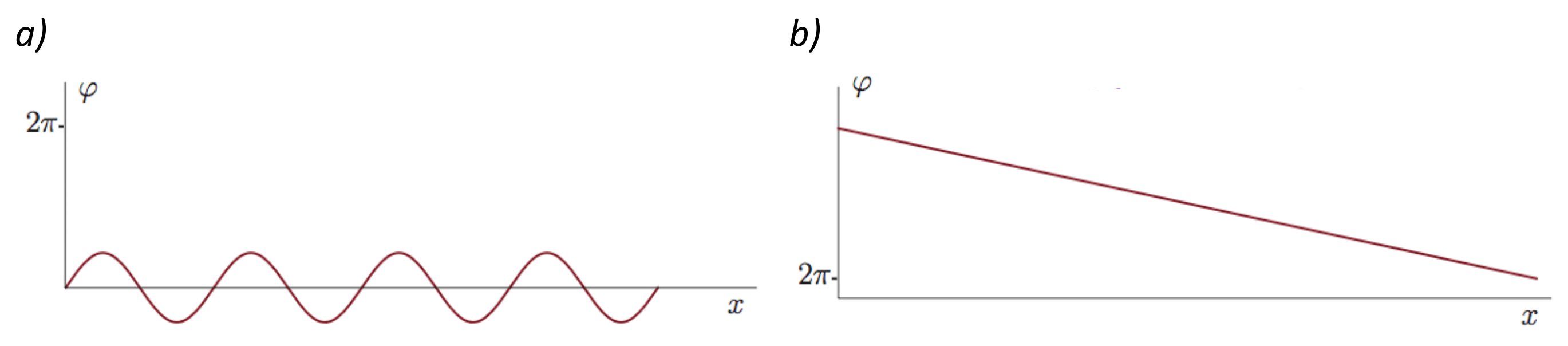}
\caption[]{ Phase (in-plane rotation angle)  variation at the presence of mass (spin) currents. a) Oscillating currents in a sound (spin) wave). b) Stationary mass (spin) supercurrent. }
\label{fig1}
\end{figure}

Since the idea of spin superfluidity emerged from the analogy of magnetodynamics and superfluid hydrodynamics let us remind the concept of the mass  superfluidity in the theory of  superfluidity. 
In superfluid hydrodynamics there are the Hamilton equations for the pair of the canonically conjugate variables ``phase -- density'':
 \begin{eqnarray}
\hbar {d\varphi \over dt}=-{\delta {\cal H}\over \delta n},~~
{dn \over dt}= {\delta{\cal H}\over \hbar \delta \varphi}.
     \label{IdHam} \end{eqnarray}
Here  
\begin{equation}
{\delta {\cal H}\over \delta n}={\partial  {\cal H}\over \partial n} -\bm \nabla \cdot {\partial  {\cal H}\over \partial \bm \nabla n},~~{ \delta {\cal H}\over \delta \varphi}={\partial  {\cal H}\over \partial \varphi} -\bm \nabla \cdot {\partial  {\cal H}\over \partial \bm \nabla \varphi} 
         \end{equation}
are functional derivatives of the Hamiltonian 
\be
{\cal H} ={\hbar^2n \over 2m}\nabla \varphi^2+ E_0(n),
    \ee{}
where $E_0(n)$ is the energy of the superfluid at rest, which depends only on the particle density $n$. Taking into account  the gauge invariance (the energy does not depend on the phase directly, $\partial {\cal H} / \partial \varphi=0$, but only on its gradient)  the Hamilton equations are reduced to  the equations of hydrodynamics for an ideal liquid:
  \begin{eqnarray}
m{d\bm v \over dt}= -\bm \nabla \mu ,
\label{Eul}           \end{eqnarray}
 \begin{eqnarray}
{dn \over dt}=-\bm \nabla \cdot \bm j.
     \label{IdLiq} \end{eqnarray}
In these expressions 
\be
\mu= {\partial E_0\over \partial n}+{\hbar^2 \over 2m}\nabla \varphi^2
    \ee{}
is the chemical potential, and
\begin{equation}
 \bm j=n\bm v ={\partial  {\cal H}\over \hbar \partial \bm \nabla \varphi} 
 \end{equation} 
 is the particle current. We consider the zero-temperature limit, when the superfluid velocity coincides with the center-of-mass velocity
 \be
 \bm v= {\hbar \over m} \bm \nabla \varphi.
          \ee{}

A collective mode of the ideal liquid is a sound wave. In the sound wave the phase varies in space, i.e., the wave is accompanied by mass currents [Fig.~\ref{fig1}(a)]. An amplitude of the phase variation is small, and currents transport mass on distances of the order of the wavelength. A real superfluid transport on macroscopic distances is possible in current states, which are stationary solutions of the hydrodynamic equations with finite constant currents, i.e., with constant nonzero phase gradients. In the current state the phase rotates through a large number of full 2$\pi$-rotations along streamlines of the current [Fig.~\ref{fig1}(b)]. These are supercurrents or persistent currents.  

\begin{figure}[t]
\includegraphics[width=.7\textwidth]{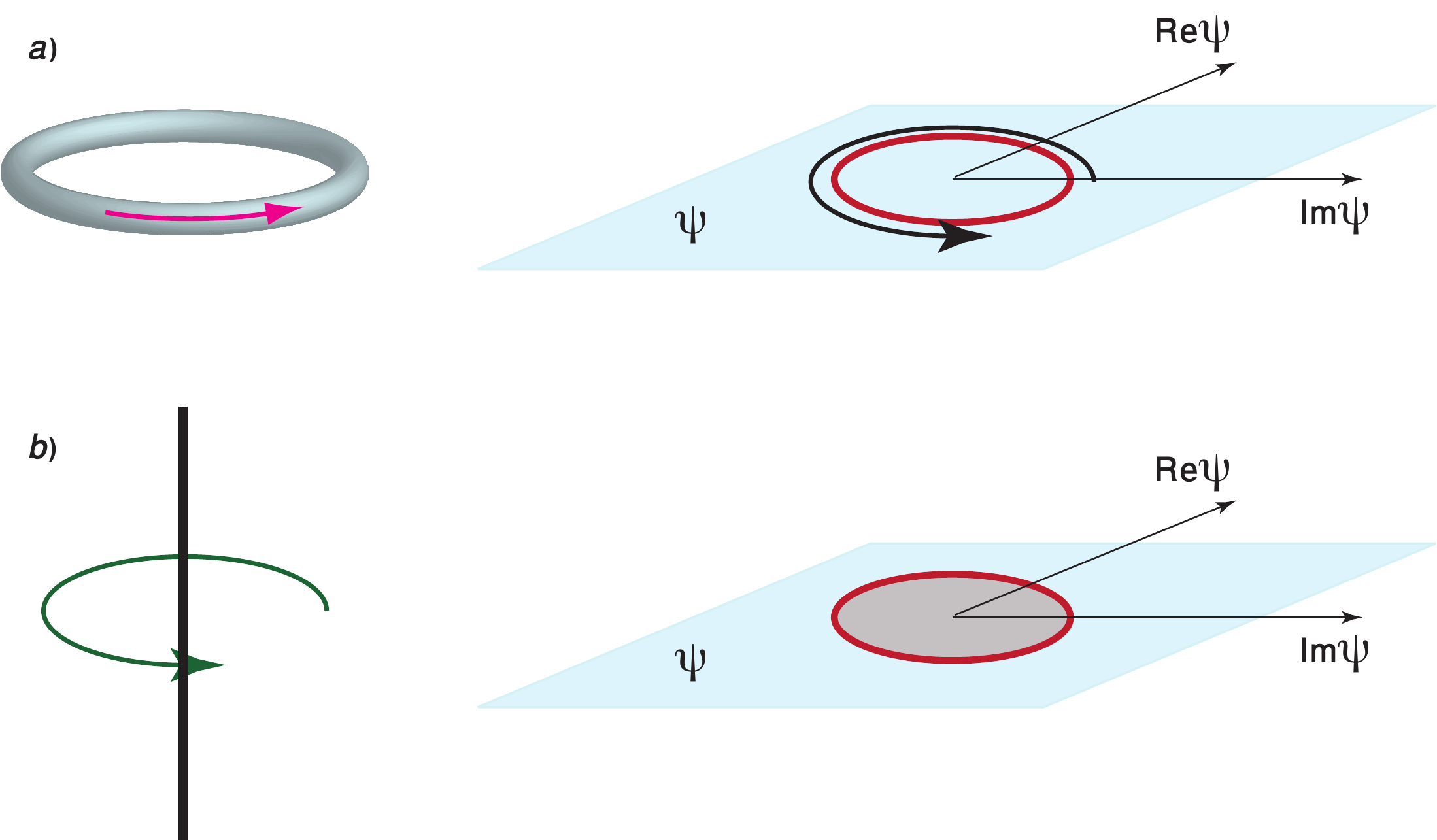}
\caption[]{ Topology of the uniform mass current and the vortex states. a) The current state in a torus maps onto the circumference $|\psi|=|\psi_0|=const$ in the complex $\psi$ - plane, where $\psi_0$ is the equilibrium order parameter wave function of the uniform state. b) The vortex state maps onto the circle $|\psi|\leq |\psi_0| $.  }
\label{fig2a}
\end{figure}

The crucial point of the superfluidity concept  is the question why the supercurrent is a persistent current, which does not decay despite it is not the ground state of the system. The answer to this question follows  from the analysis of the topology of the order parameter space (vacuum manifold).  At the equilibrium the order parameter of a superfluid is a complex wave function   $\psi = \psi_0 e^{i\varphi}$, where the modulus $\psi_0$ of the wave function is a positive constant determined by minimization of the energy and the
 phase $\varphi$ is a degeneracy parameter since the energy does not depend on  $\varphi$. Any from the degenerate ground states in a closed annular channel (torus) maps on some point at the circumference $|\psi|=\psi_0$ in the complex plane $\psi$, while a
 current state with  the phase change $2\pi n$ around the torus maps onto a path [Fig.~\ref{fig2a}(a)] winding around the  circumference $n$ times. It is impossible to change the winding number  $n$ keeping the path on the circumference $|\psi|=\psi_0$ all the time. In the language of topology states with different $n$ belong to different classes, and  $n$ is a {\em topological charge}. Only a vortex moving across the torus channel  can change $n$ to $n-1$. This process is a phase slip.  The phase slip costs energy, which is spent on creation of the vortex and its motion  across current streamlines.  The state with the vortex in the channel maps  on the full circle $|\psi| \leq \psi_0$ [Fig.~\ref{fig2a}(b)]. Thus, phase slips are  impeded by potential barriers, which make the current state metastable.
 
According to the Landau criterion, the current state is metastable as long as \emph{any quasiparticle} of the superfluid in the laboratory frame has a positive energy and therefore its creation requires an energy input.  The Landau criterion checks the stability only with respect to weak elementary perturbations of the current state, while a vortex is a strong macroscopic perturbation. However, the  Landau critical gradients are of the same order as the gradients at which barriers for phase slips disappear.  The both are on the order of the inverse vortex core radius.  

\section{Spin superfluidity}\label{magn}

\subsection{Ferromagnets}

The phenomenological description of magnetically ordered media is given by the LLG theory. For a ferromagnet with magnetization density $\bm M$ the LLG equation is \cite{LL}
\begin{eqnarray}
{\partial \bm M\over \partial t}=\gamma \left[\bm H _{eff} \times  \bm M\right],
     \label{LLP}      \end{eqnarray}
where $\gamma$ is the gyromagnetic ratio between the magnetic and mechanical moment. The effective magnetic field is determined by the functional derivative of the total energy:
\begin{eqnarray}
\bm H _{eff} =- {\delta {\cal H }\over \delta \bm M}=- {\partial {\cal H }\over \partial \bm M}+\nabla_i{\partial {\cal H }\over \partial \nabla_i\bm M}.
           \end{eqnarray}
According to the   LLG equation, the absolute value $M$ of the magnetization cannot vary.  The evolution of  $\bm M$  is a precession around the effective magnetic field $\bm H _{eff}$.        

We shall consider the case  when spin-rotational invariance is partially broken, and  there is uniaxial crystal magnetic anisotropy. The phenomenological Hamiltonian is
\begin{eqnarray}
{\cal H}= {A\over 2} \nabla_i \bm M \cdot \nabla_i \bm M+{G M_z^2\over 2M^2} - \bm H \cdot \bm M.
           \end{eqnarray}
If the anisotropy energy $G$ is positive, it is the ``easy plane'' anisotropy, which keeps the magnetization  in the $xy$ plane. If the external magnetic field $\bm H$ is directed along the $z$ axis, the $z$ component of spin is conserved because of  invariance with respect to rotations around the $z$ axis. Since the absolute value $M$ of magnetization is fixed, the magnetization vector $\bm M$  is fully determined by the $z$ magnetization component $M_z$ and the angle $\varphi $ showing the direction of $\bm M$ in the easy plane $xy$:
\be
M_x =M_\perp \cos \varphi,~~M_y =M_\perp \sin \varphi,~~M_\perp=\sqrt{M^2-M_z^2} .
    \ee{}
In the new variables the Hamiltonian is
\begin{eqnarray}
{\cal H}= {AM_\perp^2(\bm \nabla \varphi)^2\over 2}+{ M_z^2\over 2\chi}- HM_z  .
  \label{Ener}    \end{eqnarray}
Here we neglected gradients of $M_z$. The parameter  $A$ is stiffness of the spin system determined by exchange interaction, and
the magnetic susceptibility $\chi= M^2/G$ along the $z$ axis is determined by the uniaxial anisotropy energy $G$ keeping the magnetization in the easy plane.  The LLG equation reduces to the Hamilton equations for a pair of canonically conjugate continuous variables ``angle--angular momentum'':
   \begin{eqnarray}
{1\over \gamma}{d\varphi \over dt}=- {\delta {\cal H} \over \delta M_z}=- {\partial {\cal H} \over \partial M_z},
     \label{HEp} \end{eqnarray}
     \begin{eqnarray}
{1\over \gamma}{dM_z \over dt}={\delta {\cal H}\over \delta \varphi}=-\bm \nabla \cdot{\partial {\cal H} \over \partial \bm \nabla \varphi},
 \label{HEm}      \end{eqnarray}
where functional derivatives on the right-hand sides are taken from the Hamiltonian  given by Eq.~(\ref{Ener}). Using the expressions for functional derivatives one can write the Hamilton equations as
         \begin{eqnarray}
{1\over \gamma}{d\varphi \over dt}=AM_z(\bm \nabla \varphi)^2-{ M_z-\chi  H\over \chi},
     \label{Ep} \end{eqnarray}
     \begin{eqnarray}
{1\over \gamma} {dM_z \over dt}+ \bm \nabla \cdot \bm J =0,
 \label{Em}      \end{eqnarray}
where 
\begin{eqnarray}
\bm J=-{\partial {\cal H} \over \partial \bm \nabla \varphi} =-A M_\perp^2  \bm \nabla \varphi
   \label{cur}    \end{eqnarray}
is the spin current. Although our  equations contain not the spin density but the magnetization, the vector $\bm J$ is defined as a current of spin with the spin density $M_z/\gamma$.

\begin{figure}[b]
\includegraphics[width=.75\textwidth]{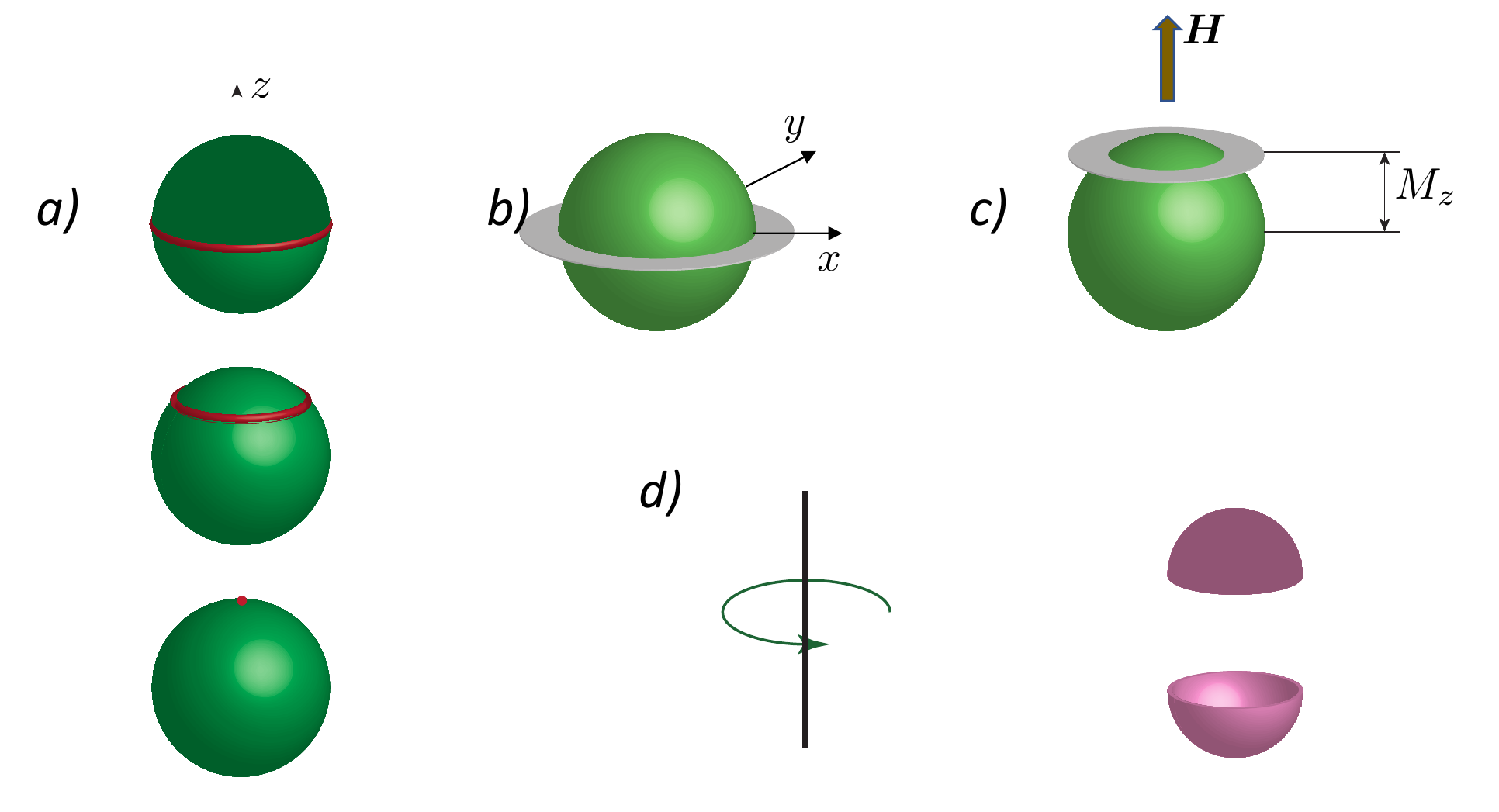}
\caption[]{Mapping of spin current states on the order parameter space (vacuum manifold).\newline
a) Spin currents in an isotropic ferromagnet. The current state in torus maps on an equatorial circumference on the sphere of radius $M$ (top). Continuous shift of mapping on the surface of the sphere (middle) reduces it to a point at the northern pole (bottom), which corresponds to the ground state without currents. \newline
b) Spin currents in an easy-plane ferromagnet. The easy-plane anisotropy reduces the order parameter space to an equatorial circumference in the $xy$ plane topologically equivalent to the order parameter space in superfluids. \newline
c) Spin currents in an easy-plane ferromagnet in a magnetic field parallel to the axis $z$. Spin is confined in  the plane parallel to the $xy$ plane but shifted  closer to the northern pole.  \newline 
d) The vortex state maps on the surface of  the upper or the lower  semisphere in the vacuum manifold.}
\label{Fig02}
\end{figure}

There is an evident analogy of Eqs.~(\ref{Ep}) and (\ref{Em}) with the hydrodynamic equations (\ref{Eul})  and (\ref{IdLiq})  for the superfluid. One of  solutions of these equations describes  the spin-wave mode. However, as well as the mass current in a sound wave, the small oscillating spin current in the spin wave does not lead to long-distance superfluid spin transport, which this review addresses. Spin superfluid transport on long distances is realized in current states with spin rotating in the plane through a large number of full 2$\pi$-rotations as shown in Fig.~\ref{fig1}(b). In the current state with a constant gradient of the spin phase $\bm K=\bm \nabla \varphi$, there is a constant magnetization component along the magnetic field (the axis $z$):
\be
M_z={\chi H\over 1-\chi AK^2}.
      \ee{Fmag}

 Like in superfluids, the stability of current states is connected to the topology of the order parameter space. 
In isotropic ferromagnets ($G=0$) the order parameter space is a spherical surface of radius equal to the absolute value of the magnetization vector $\bm M$ [Fig.~\ref{Fig02}(a)]. All points on this surface correspond  to the same energy of the ground state.  Suppose we created the spin current state with monotonously varying  phase $\varphi$ in a torus. This state maps on the equatorial circumference in the order parameter space. The topology allows to continuously shift the circumference and to reduce it to a point  (the northern or the southern pole).  During this process shown in Fig.~\ref{Fig02}(a)  the path remains in the order parameter space all the time, and therefore, no energetic barrier resists to the transformation. Thus, the metastability of the current state is not expected in isotropic ferromagnets.

In a ferromagnet with easy-plane anisotropy ($G>0$) the order parameter space reduces from the spherical surface to the equatorial circumference in the $xy$ plane [Fig.~\ref{Fig02}(b)]. This makes the order parameter space topologically equivalent to that in superfluids.  Now the transformation of the circumference to the point   costs the anisotropy energy. This allows to expect metastable spin currents (supercurrents). The magnetic field along the anisotropy axis $z$ shifts the easy plane either up [Fig.~\ref{Fig02}(c)] or down away from the equator.

The current states in easy-plane ferromagnets relax to the ground state via phase slips events, in which magnetic vortices cross spin current streamlines.  States with vortices map on  a surface of a hemisphere of radius $M$ either above or below the equator\cite{Nik} as shown in Fig.~\ref{Fig02}(d).

Up to now we considered states close to the equilibrium (ground) state. In a ferromagnet  in a magnetic field the equilibrium magnetization is parallel to the field. However,  by pumping magnons into the sample it is possible to tilt the magnetization with respect to the magnetic field. This creates the state with the coherent spin precession around the magnetic field (the magnon BEC\cite{BunV,Dem6}). 
Although the state is far from the true equilibrium, but it, nevertheless, is a state of minimal energy at fixed magnetization $M_z$. Because of inevitable spin relaxation  the state of uniform precession requires permanent pumping of spin and energy. However, if  processes violating the spin conservation law are weak, one can ignore them and treat the state as a quasi-equilibrium state. The state of uniform precession maps on a circumference parallel to the $xy$ plane.   One can consider also a current  state, in which the phase (the rotation angle in the $xy$ plane) varies not only in time but also in space with a constant gradient. In this case the easy plane for the magnetization is not related to the crystal anisotropy but created dynamically. However, in the quasi-equilibrium coherent precession state demonstration of the long-distance  superfluid spin transport is problematic (see Sec.~\ref{DC}).

\subsection{Antiferromagnets}

Long time ago it was widely accepted  to describe the dynamics of a bipartite antiferromagnet by the LLG equations for two spin sublattices coupled via exchange interaction:\cite{Kittel51} 
\be
{d\bm M_i\over dt}=\gamma \left[\bm H_i  \times \bm M_i\right].
       \ee{LLG}
 Here the subscript  $i=1,2$ indicates to which sublattice the magnetization $\bm M_i$ belongs,   and
 \be
\bm H_i =-{\delta {\cal H}\over \delta \bm M_i}= - {\partial {\cal H}\over \partial \bm M_i}+\nabla_j{\partial {\cal H}\over \partial \nabla_j\bm M_i} 
   \ee{}     
is the effective field for  the $i$th  sublattice determined by the functional derivative of the Hamiltonian $\cal H$.  For an isotropic antiferromagnet  the Hamiltonian is 
\be
{\cal H}= {\bm M_1 \cdot \bm M_2\over \chi} + { A (\nabla_i \bm M_1 \cdot \nabla_i \bm M_1+\nabla_i \bm M_2 \cdot \nabla_i \bm M_2)\over 2}
+A_{12} \nabla_j \bm M_1 \cdot \nabla_j \bm M_2-\bm H \cdot \bm m.
  \ee{ham2}
In the uniform ground state the total magnetization
\be 
\bm m=\bm M_1+\bm M_2
 \ee{}
 is equal to $\bm m =\chi \bm H$,  while the staggered magnetization
 \be
 \bm L=\bm M_1-\bm M_2
   \ee{}
is normal to $\bm m$. Without the magnetic field   the two sublattice magnetizations are antiparallel, and the total magnetization $\bm m$ vanishes.  The first term in the Hamiltonian (\ref{ham2}), which determines the susceptibility $\chi$,  originates from the exchange interaction between spins of the two sublattices. This is the susceptibility normal to the staggered magnetization $\bm L$. Since in the LLG theory absolute values of sublattice magnetizations $\bm M_1$ and $\bm M_2$ are equal to $M$ and do not vary in space and time, the susceptibility parallel to $\bm L $ vanishes.

Let us consider a uniform state but not necessary the ground state. There are no  currents in this state, and the gradient-dependent terms in the Hamiltonian \eq{ham2} can be ignored. Rewriting the Hamiltonian in terms of $\bm m$ and $\bm L$ one obtains
\be
{\cal H}= -{L^2-m^2 \over  4 \chi} -\bm H \cdot \bm m = -{M^2 \over   \chi}+{m^2 \over  2 \chi} - \bm H \cdot \bm m.
  \ee{ham2r}
 Minimizing the Hamiltonian with respect to the  absolute value of $\bm m$ (at it fixed direction) one obtains
\bem
{\cal H}=- {M^2 \over   \chi}-{\chi  H_m^2 \over  2 }=- {M^2 \over   \chi}-{\chi  H^2 \over  2 } +{\chi  H_L^2 \over  2},
  \eem{HML}
where  $H_m= (\bm H \cdot \bm m)/m$ and $H_L =(\bm H \cdot \bm L)/L$  are the projections of the magnetic field on the total magnetization $\bm m$ and on the staggered magnetization $\bm L$.  The first two terms are constant, while the last term plays the role of the easy-plane anisotropy energy confining $\bm L$ in the plane normal to $\bm H$. For $\bm H$ parallel to the axis $z$ the anisotropy energy [the last term on the right-hand side of \eq{HML}] is
\be
E_a = {\chi  H^2 L_z^2 \over 2 L^2}.
        \ee{Ea}

In the analogy to the ferromagnetic case, one can describe the vectors of sublattice magnetizations $\bm M_i$ with the constant absolute  value $M$ by the two pairs of the conjugate variables $(M_{iz},\varphi_i)$, which are determined by the two pairs of the Hamilton equations:
 \begin{eqnarray}
{1\over \gamma}{d\varphi_i \over dt}=- {\delta {\cal H} \over \delta M_{iz}}=- {\partial {\cal H} \over \partial M_{iz}},
     \label{aEp} \end{eqnarray}
     \begin{eqnarray}
{1\over \gamma}{dM_{iz} \over dt}={\delta {\cal H}\over \delta \varphi_i}={\partial {\cal H}\over \partial \varphi_i}-\bm \nabla \cdot{\partial {\cal H} \over \partial \bm \nabla \varphi_i}.
 \label{aEm}      \end{eqnarray}
Let us consider the axisymmetric solutions of these equations with $\varphi=\varphi_1=\pi-\varphi_2$ and $M_{1z}=M_{2z}={m_z\over 2}$. Then there is only one pair of the Hamilton equations for the pair of the conjugate variables $(m_z,\varphi)$:
        \begin{eqnarray}
{1\over \gamma}{d\varphi \over dt}={A_-m_z(\bm \nabla \varphi)^2\over 2}-{ m_z-\chi  H\over \chi},
     \label{AFEp} \end{eqnarray}
     \begin{eqnarray}
{1\over \gamma} {dm_z \over dt}+ \bm \nabla \cdot \bm J =0.
 \label{AFEm}      \end{eqnarray}
Here 
\begin{eqnarray}
\bm J=-{\partial {\cal H} \over \partial \bm \nabla \varphi} =-{A_- L^2\over 2}  \bm \nabla \varphi
   \label{AFcur}    \end{eqnarray}
is the the spin current
and $A_-=A-A_{12}$. These equations are identical to Eqs.~(\ref{Ep}) and (\ref{Em}) for the ferromagnetic after replacing the spontaneous magnetization component $M_z$ by the total magnetization component $m_z$, $A$ by $A_-/2$, and $M_\perp $ by $L$. In the stationary current  state there is a constant gradient $\bm K=\bm \nabla\varphi$ of the spin  phase and a constant total magnetization
\be
m_z = {\chi H \over 1-\chi A_- K^2/2}.
   \ee{mzA}
While in ferromagnets the current state is a spiral spin structure with the spatial precession of the in-plane spontaneous magnetization along current streamlines, in antiferromagnets the current states are related to the spatial precession  of the staggered magnetization.

The order parameter space for the isotropic antiferromagnet in the absence of the external magnetic field is a surface of a sphere. However, the order parameter is not the total magnetization but   the unit N\'eel vector $\bm l=\bm L/L$. While in the ferromagnet the magnetic field produces an easy axis for the total magnetization, in the antiferromagnet the magnetic field produces the easy plane for the order parameter vector $\bm l$  with the anisotropy energy given by \eq{Ea}. Thus, the topology necessary for the spin superfluidity in antiferromagnets does not require the crystal easy-plane anisotropy.  

\section{Spin currents without spin conservation law}    \label{Phase}

Though processes violating the spin conservation law  are relativistically weak, their  effect is of principal importance and cannot be ignored in general. The attention to superfluid transport   in the absence of conservation law was attracted first in discussions of superfluidity of electron-hole pairs. The number of electron-hole pairs can vary due to interband transitions, and  the degeneracy with respect to the phase of the pair condensate is lifted.  On the basis of it Guseinov and Keldysh\cite{GuKe} concluded that the existence of spatially \emph{homogeneous} stationary current states is impossible and there is no analogy with superfluidity.  
This phenomenon was called ``fixation of phase''. However some time  later it was demonstrated \cite{ES-77} that phase fixation does not rule out the existence of weakly \emph{inhomogeneous} stationary current states analogous to superfluid current states.\footnote
{Similar conclusions have been done with respect to possibility of supercurrents in systems with spatially separated electrons and holes.\cite{Sh,LY}} This analysis was extended on spin superfluidity.\cite{ES-78a,ES-78b}

\begin{figure}[b]
\includegraphics[width=.9\textwidth]{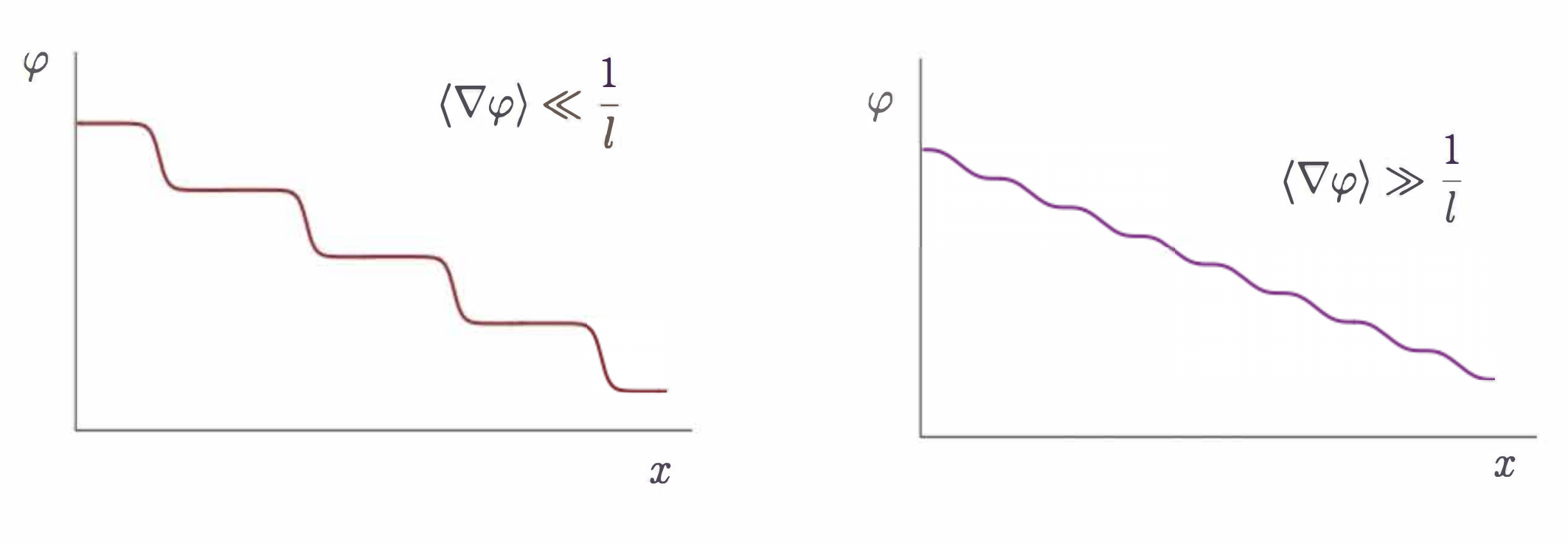}
\caption[]{ The nonuniform spin-current states with $\langle \nabla \varphi \rangle \ll 1/l$ and $\langle \nabla \varphi \rangle \gg 1/l$. }
\label{fig2}
\end{figure}

One can take into account processes violating the spin conservation law by adding the $n$-fold in-plane anisotropy energy $\propto G_{in}$ to the Hamiltonian (\ref{Ener}):
\begin{eqnarray}
{\cal H}={ M_z^2\over 2\chi}-\gamma M_z H+ {AM_\perp^2(\bm \nabla \varphi)^2\over 2}+G_{in}[1-\cos (n\varphi)] .
 \label{an}  \end{eqnarray}
Then the spin continuity equation (\ref{Em}) becomes
   \begin{eqnarray}
{dM_z \over dt}=-\bm \nabla \cdot \bm J+nG_{in} \sin(n\varphi)= AM_\perp^2\left[\nabla^2 \varphi -{\sin(n\varphi)\over l^2}\right] ,
 \label{EmK}      \end{eqnarray}
where 
 \begin{eqnarray}
l=\sqrt{ AM_\perp^2\over nG_{in}}
             \end{eqnarray}
is the thickness of the wall separating domains with  $n$ equivalent easiest directions in the easy plane.
We focus on stationary states when $dM_z/dt=0$.  The phase $\varphi$ is a periodical solution of the sine-Gordon equation parametrized by the average phase gradients $\langle \nabla \varphi \rangle$. At small $\langle \nabla \varphi \rangle \ll 1/l$ the spin structure constitutes the chain of domains with the period $2\pi/n\langle \nabla \varphi \rangle$. Any  domain corresponds  to some of the $n$ equivalent easiest directions in the easy plane.  Spin currents (gradients) inside domains are negligible but there are essential spin currents inside domain walls where  $\nabla \varphi  \sim 1/l$. This hardly reminds the  superfluid transport on macroscopic scales: spin is transported over distances on the order of the domain-wall thickness $l$. With increasing $\langle \nabla \varphi \rangle$ the density of domain walls grows, and at $\langle \nabla \varphi \rangle\gg 1/l$ the domains coalesce. Deviations of the gradient $\nabla \varphi$ from the constant average gradient $\langle \nabla \varphi\rangle$  become negligible. This restores the analogy with the superfluid transport in superfluids. The transformation of the domain wall chain into a weakly inhomogeneous current state at growing $\langle \nabla \varphi \rangle$  is illustrated in Fig.~\ref{fig2}. 

An important difference with conventional mass superfluidity is that the existence of conventional superfluidity is restricted only from above by the Landau critical gradients, while  the existence of spin superfluidity is restricted also from below: gradients should not  be less than the value $1/l$. Since the upper Landau critical value is determined by the easy-plane uniaxial anisotropy $G$  and the lower critical value is determined by the in-plane anisotropy energy $G_{in}$, spin superfluidity is possible only if $G\gg G_{in}$. The existence of the lower critical gradient for spin superfluidity  is important for spin superfluidity observation discussed in Sec.~\ref{disc}. However, in the further theoretical analysis we ignore processes violating the spin conservation law assuming that the phase gradients essentially exceed the lower threshold for spin superfluidity.

\section{Collective spin modes and the Landau criterion}

\subsection{Ferromagnets}

In order to check the Landau criterion, one should know the spectrum of collective modes in the current state with the constant value of the spin phase gradient $\bm K=\bm  \nabla \varphi$ and with the longitudinal (along the magnetic field) magnetization  given by \eq{Fmag}.
It is necessary to solve the Hamilton equations Eqs.~(\ref{Ep}) and  (\ref{Em}) linearized with respect to
weak perturbations of the current state.  We skip the standard algebra given elsewhere\cite{Son19}. Finally one obtains\cite{Hoefer,Son19} the spectrum of plane spin waves $\propto e^{i\bm k\cdot \bm r-i\omega t}$:
\be
\omega +\bm w \cdot \bm k = \tilde c_{sw}  k.
    \ee{spF}
Here
\be
\tilde c_{sw}  = \sqrt{ 1 - \chi A K^2} c_{sw} 
      \ee{}
is the spin-wave velocity in the current state and  
\be
c_{sw}  =\gamma M_\perp \sqrt{A\over \chi}
    \ee{}
is the spin velocity in the state without spin currents.
The velocity
\be
\bm w =2\gamma M_z A \bm K
     \ee{}    
can be called Doppler velocity because its effect  on the mode frequency is similar to the effect of the mass velocity on the mode frequency in a Galilean invariant fluid (Doppler effect). However, our system is not Galilean invariant,\cite{Hoefer} and the gradient $K$ is present also on the right-hand side of the dispersion relation (\ref{spF}). 

We obtained the gapless Goldstone mode with the sound-like  linear in $\bm k$ spectrum.  The current state becomes unstable   when  at $ \bm k$ parallel to $\bm w$ the frequency $\omega$ becomes negative. This happens at the gradient  $K$ equal to the Landau critical gradient
\be
K_c={M_\perp \over\sqrt{ 4M^2 - 3M_\perp}} {1\over \sqrt{\chi A}}.
     \ee{KcFg}
In the limit of weak magnetic fields when $M_z\ll M$ the Landau critical gradient is
\be
K_c= {1\over \sqrt{\chi A}}= {\gamma  M \over \chi c_{sw} }.
     \ee{KcF}
In this limit the  pseudo-Doppler effect is not important, and the Landau critical gradient $K_c$ is determined by the condition that the spin-wave velocity $\tilde c_{sw} $ in the current state vanishes.

In the opposite limit $M_z \to M$ ($M_\perp \to 0$) the Landau critical gradient, 
\be
K_c={M_\perp \over 2M} {1\over \sqrt{\chi A}},
     \ee{}
decreases, and the spin superfluidity becomes impossible at the phase transition to the easy-axis anisotropy  ($M_\perp=0$). 

Deriving the sound-like spectrum of the spin wave we neglected in the Hamiltonian   terms dependent on gradients  $\bm \nabla M_z$. Taking these terms into account one obtains quadratic in $k$ corrections to the spectrum. These corrections  become important at $k \sim 1/\xi_0$, where 
\be
\xi_0={M\over M_\perp}\sqrt{ \chi A}
    \ee{xi0}
can be called the coherence length. The coherence length $\xi_0$ determines the core radius of vortices just because the gradients 
$\bm \nabla M_z$ are important in the vortex core. On the other hand, the calculation of the energy of the vortex in the current state (Sec.~\ref{PS}) indicates that potential barriers for phase slips disappear at the gradients of the order $1/\xi_0$. Since the $1/\xi_0$ is of the same order of magnitude as the Landau critical gradient \eq{KcFg}, the instability with respect to elementary excitations (the Landau instability) and the instability with respect to macroscopic  excitations (vortices participating in phase slips) start at approximately the same gradients. 

\subsection{Antiferromagnets}

Directions of the sublattice magnetizations in a bipartite antiferromagnet are determined by the two pairs of polar angles $\theta_i$, $\varphi_i$ ($i=1,2$):
\bem
M_{ix} =M\cos\theta_i\cos \varphi_i,~~M_{iy} =M\cos\theta_i\sin \varphi_i,
~~M_{iz}=M\sin \theta_i.
       \eem{}
      
\begin{figure}[b]
\includegraphics[width=.4\textwidth]{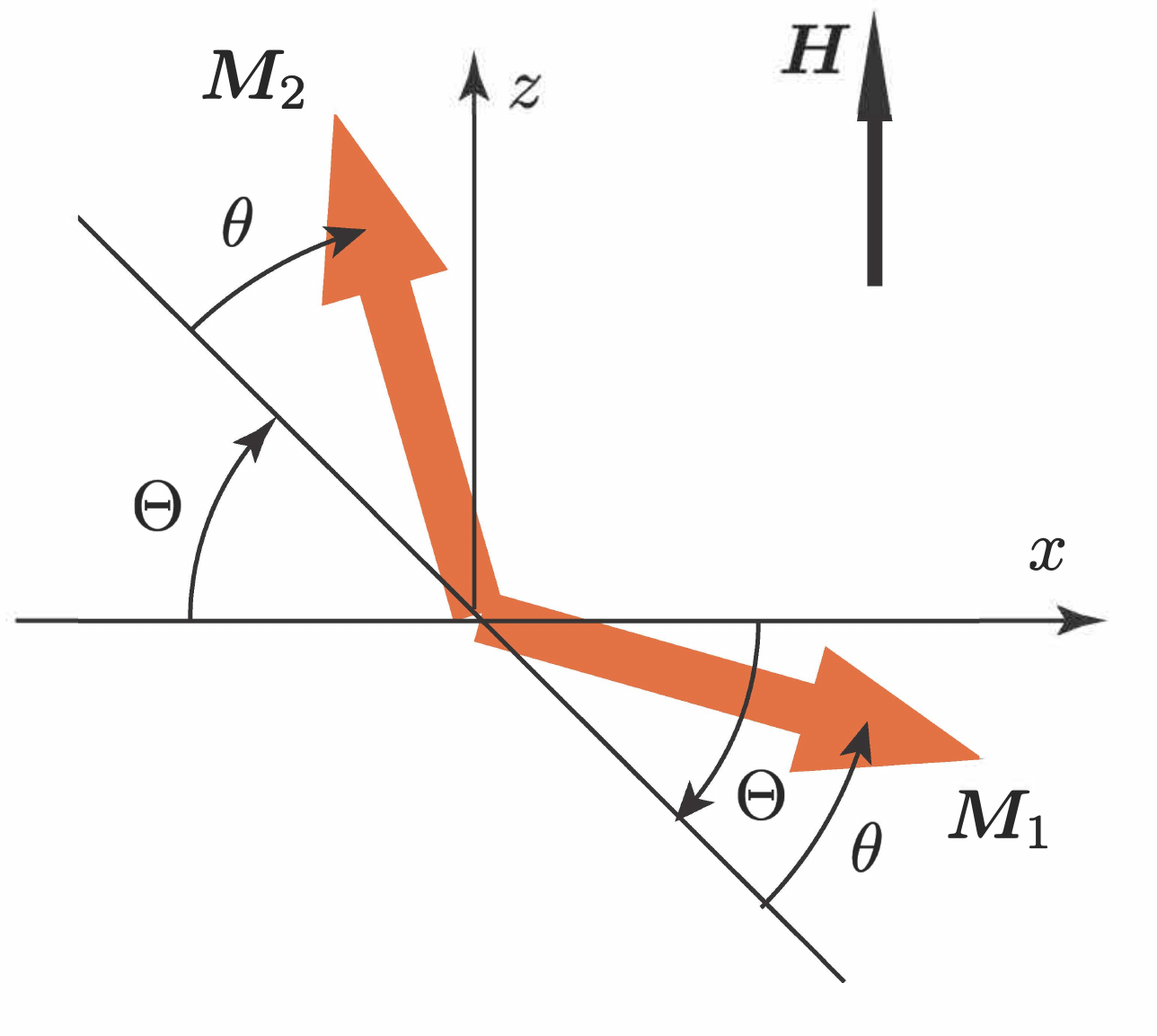}
\caption[]{ Angle variables $\theta$ and $\Theta$  for the case when the both magnetizations are in the plane $xz$ ($\varphi =\Phi=0$).}
\label{f2}
\end{figure}

 In the further analysis it is convenient to use other angle variables:
\bem
\theta={\pi +\theta_1-\theta_2\over 2},~~  \Theta ={\pi-\theta_1-\theta_2\over 2},
\nonumber \\
\varphi={\varphi_1+\varphi_2\over 2},~~\Phi={\varphi_1-\varphi_2\over 2}.
    \eem{}
The polar angle $\Theta$ for the staggered magnetization $\bm L$ and the canting angle $\theta$ are shown in Fig.~\ref{f2} for the case when the both magnetizations are in the plane $xz$ ($\varphi =\Phi=0$). In these angle variables the equations for two collective modes in antiferromagnets are decoupled.

 In the stationary current state the total magnetization $m_z =2M \sin\theta$ is given  by \eq{mzA} and $\bm \nabla \varphi=\bm K$, while $\Theta=\Phi=0$.  In a weakly perturbed current state  small but nonzero  $\Theta$ and $\Phi$ appear. Also the angles $\theta$ and $\varphi$ differ from their values in the stationary current state:  $\theta \to \theta+ \theta'$,   $\varphi \to\varphi+ \varphi'$. As in the ferromagnetic case, we skip the  algebra of the linearization  and the solution of linearized equations  (see Ref.~\onlinecite{Son19} for a detailed calculation) and give  the resulting spectra of two spin-wave modes.

The equations for the pair of perturbations $(\theta',\varphi')$ describe the Goldstone mode with the spectrum of the plane spin waves
\bem
\omega +\bm w\cdot \bm k
= \tilde c_{sw}.
    \eem{}
Here  the spin-wave velocity in the ground  state without spin currents, the spin-wave velocity in the current state, and  
the  Doppler velocity are given by
\be
c_{sw} =\gamma L_\perp \sqrt{ A_-\over 2 \chi},
~~\tilde c_{sw}=c_{sw}\sqrt{1 -{\chi A_-  K^2\over 2}},
   ~~\bm w=\gamma m_z A_-  \bm K,
      \ee{}
where $L_\perp =L\cos \theta$. The gapless Goldstone mode in an antiferromagnet does not differ from that in a ferromagnet, if one replaces in the expressions for the ferromagnet   $A$ by $A_- /2$ and  $M$ by $2M$.

The equations for the pair of perturbations $(\Theta,\Phi)$ describe
the gapped mode with the spectrum
\bem
\omega + \bm w\cdot \bm k=\sqrt{\omega_0^2 +c_{sw}k^2} ,
    \eem{spG}
where the gap is given by
\be
\omega_0=\sqrt{ {\gamma^2m_z ^2\over \chi^2}  - c_{sw}^2K^2}.
    \ee{sp0}

\begin{figure}[t]
\includegraphics[width=.9\textwidth]{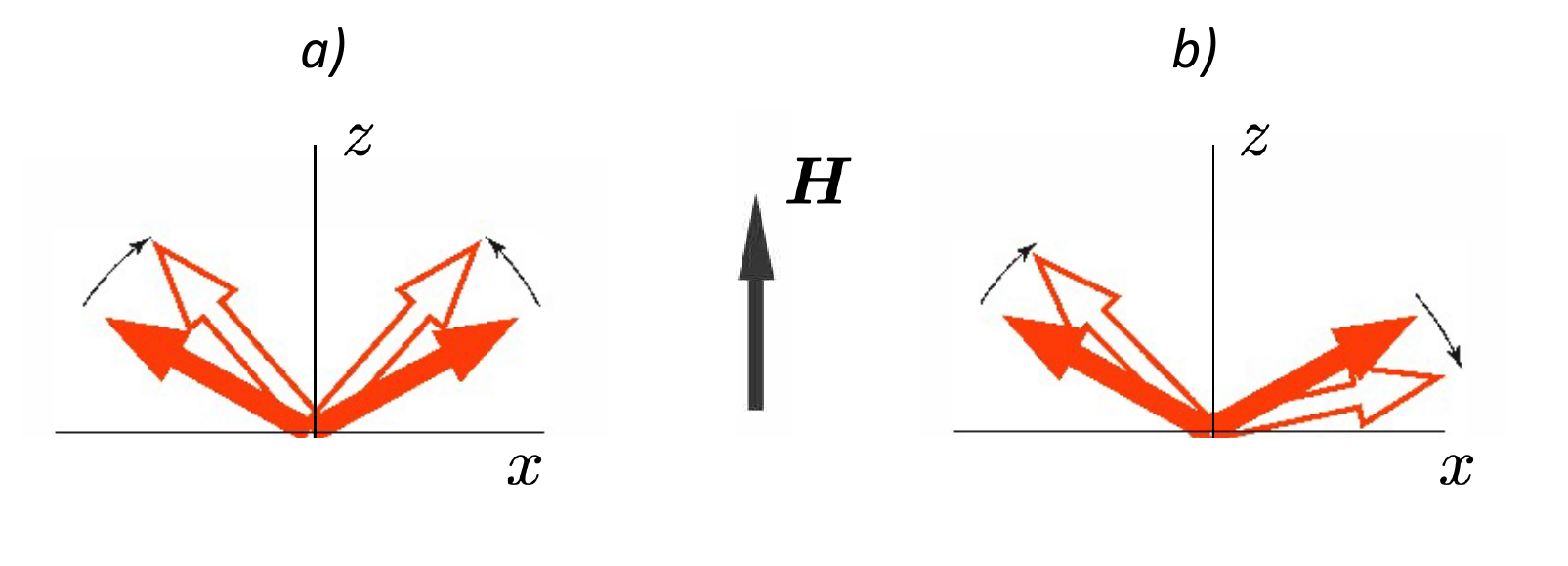}
\caption[]{ The schematic picture of the two spin wave modes in the bipartite antiferromagnet in the plane $xz$. (a) The gapless Goldstone mode. There are oscillations of the canting angle and of the total magnetization component $m_z$ and rotational oscillations around the axis $z$.
(b) The gapped mode.  There are oscillations of  the total magnetization component $m_y$ and rotational oscillations around the axis $y$. }
\label{TM}
\end{figure}

For better understanding of the physical nature of the two modes, let us consider variations of the Cartesian components of the total and the staggered  magnetizations produced by these perturbations in the uniform ground state without current and with $\bm L$ parallel to the axis $x$ ($\varphi=0$):
\bem
m'_x =2M  \sin \theta  \Theta,~~m'_y =2M \cos \theta \Phi,~~m'_z= 2M\cos \theta \theta',
\nonumber \\
L'_x =-2\sin \theta \theta',~~L'_y =2M \cos \theta  \varphi',~~L'_z= -2M \cos \theta \Theta,
     \eem{}
From these expressions one can see that the pair of perturbations $(\theta',\varphi')$ is related to the Goldstone mode and produces rotation of the staggered magnetization $\bm L$ by the angle $\varphi'$ around the axis $z$ and the oscillation of the total spin component $m'_z$  along the same axis.  On the other hand, the pair of perturbations $(\Theta,\Phi)$ produces rotation of all spins by the angle $\Theta$ around the axis $y$ and the oscillation $m'_y$ of the total spin component along the same axis. This is connected to the degree of freedom described by the pair of conjugate variables $(m_y,\Theta)$. 
In the presence of the magnetic field the rotational invariance for the axis $y$ is broken, and the mode must have a gap. In the current state with the gradient of the angle $\varphi$ the gapped mode is connected with the rotation of $\bm L$ around the axis which itself rotates in the easy-plane $xy$ along the current streamlines. Two modes are illustrated in Fig.~\ref{TM}.

In the past decades there were numerous calculations of the spin wave spectrum both in ferro- and antiferromagnets. However, the spin wave spectra discussed in the present paper were calculated not for the ground state, but for the metastable current states. In the  magnetodynamics of antiferromagnets it was usually assumed that the spin polarization is weak and the canting angle is small. Then the magnetodynamics can be reduced to the single equation for the N\'eel vector equivalent to that in the sigma model (see the recent review by \citet{Ivan18} and references therein). The derivation of spectra presented in this paper did not use the assumption of small canting angles. Therefore, the obtained dispersion relations are valid up to the magnetic field at which the sublattice magnetizations  become equal, and the staggered magnetization $\bm L$ vanishes. This makes the spin superfluidity impossible. However, this magnetic field is on the order of the exchange field, which is usually very strong.

Applying the Landau criterion to the gapless mode at small canting angles $\theta$, one obtains  the critical gradient 
\be
K_c=\sqrt{2\over \chi A_-},
   \ee{}  
similar to the value Eq.~(\ref{KcF}) obtained for a ferromagnet. However, in contrast to a ferromagnet where the susceptibility $\chi$ is connected with weak anisotropy energy, in an antiferromagnet the susceptibility $\chi$ is determined by a much larger exchange energy and is rather small. As a result, in an antiferromagnet the gapless Goldstone mode  becomes unstable at the very high value of $K$. At much lower values of $K$ the gapped mode loses its stability when the gap becomes negative and the mode frequency becomes complex. According to the spectrum (\ref{spG}), the gap in the spectrum vanishes at the critical gradient 
\be
K_c ={1\over \xi} ={\gamma H\over c_s}  ={\gamma m_z \over \chi c_s}.
   \ee{} 
Here we introduced a new correlation length
\be
\xi ={c_s\over \gamma H},
    \ee{xi}
which is connected to the effective easy-plane anisotropy energy (\ref{Ea}). The instability of the gapped mode is a precursor of the instability with respect to phase slips with vortices, which have the core radius of the order of $\xi$.

\section{Phase slips and barriers for  vortex motion across streamlines} \label{PS}

\begin{figure}[b]
\includegraphics[width=.7\textwidth]{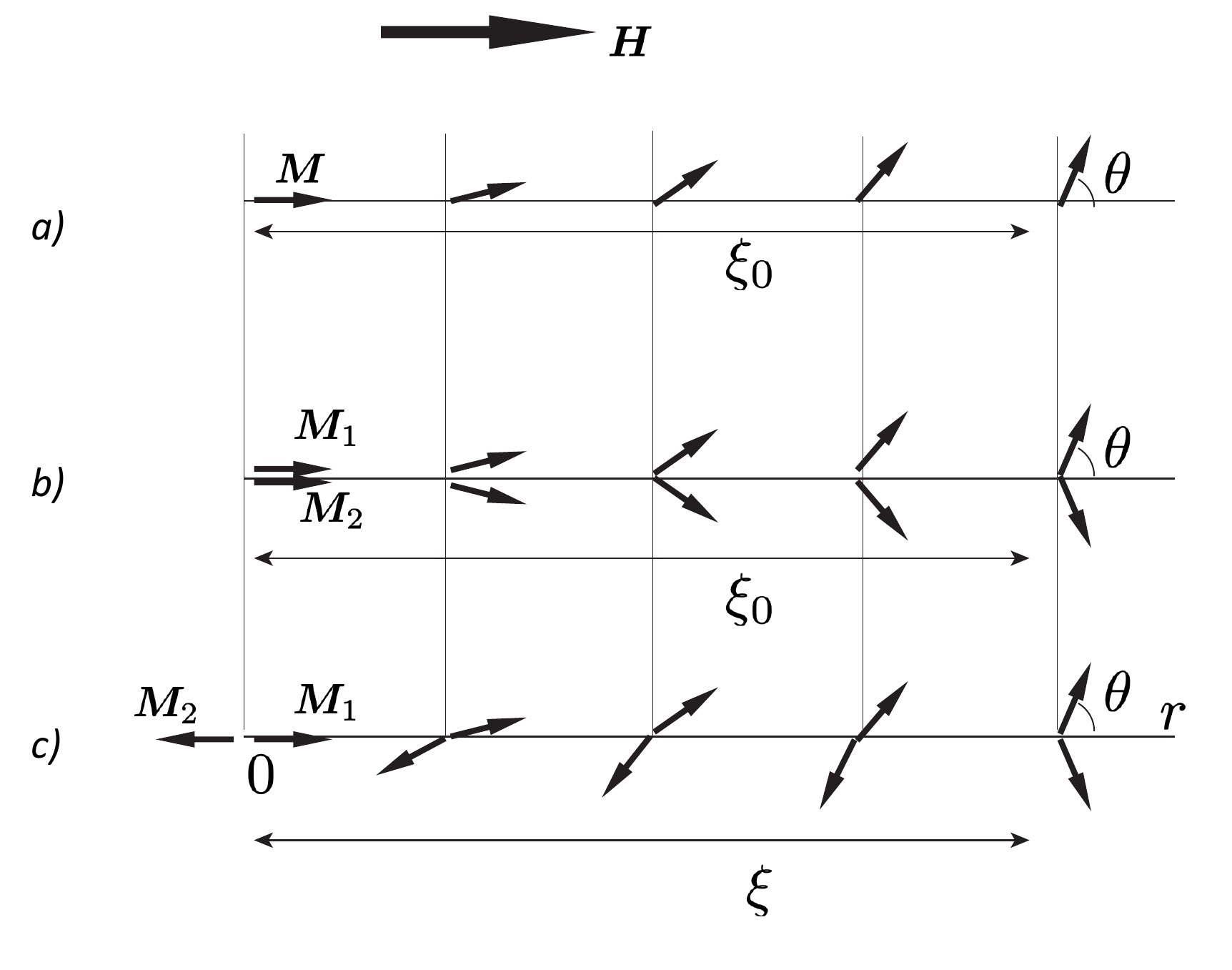}
\caption[]{ Skyrmion cores of vortices. Variation of magnetization vectors ($\bm M$ in a ferromagnet,  $\bm M_1$ and $\bm M_2$ in an antiferromagnet) in the vortex core as a function of the distance $r$ from the vortex axis is shown schematically. (a) The vortex in the ferromagnet corresponding to the single spin wave mode with the coherence length $\xi_0$ given by \eq{xi0}. 
(b) The vortex in the antiferromagnet corresponding to the Goldstone mode spin wave mode with the coherence length  $\xi_0$ given by \eq{xi0}, where the order parameter stiffness $A$ in the ferromagnet is replaced by  the stiffness $A_-/2$ in the antiferromagnet.   
(c) The vortex in the antiferromagnet corresponding to the gapped mode with the coherence length $\xi $ given by \eq{xi}. }
\label{SC}
\end{figure}

For the estimation of barriers for phase slips one must consider the interaction of vortices with spin currents. 
The total energy of the vortex is mostly determined by the area outside the core (the London region) where one must take into account  the interaction of vortices with  spin currents. In the London region the main contribution to the energy is the first term in the Hamiltonian \eq{Ener} proportional to $\nabla \varphi^2$ (we consider now a ferromagnet). This term plays the role of the kinetic energy of spin currents. The other terms are constants.  The spin phase gradient in the  current state with a straight vortex parallel to the axis $z$  is 
\be
\bm \nabla \varphi  ={[ \hat z \times \bm r]\over r^2}+\bm K,
    \ee{}
where the first term is the spin  phase gradient introduced by the vortex, $\hat z$ is the unit vector along the $z$ axis,  $\bm r$ is a 2D position vector with the origin at the vortex axis, and the gradient $\bm K$ is related to the spin current. Substituting this into the kinetic energy and integrating the energy over the whole space occupied by the ferromagnet  one obtains a logarithmically divergent integral, which depends on the sample geometry. We consider a 2D problem of the straight vortex at the distance $R$ from the plane border. The gradient $\bm K$ is parallel to the border.
Then the  the energy of the straight vortex per unit length in the presence of currents is
\bem
E_v=\pi AM_\perp^2\left(\ln {R\over r_c} - 2KR\right). 
   \eem{Ven}
The lower cutoff of the logarithm is the core radius $r_c$. The vortex energy has a maximum at $R =1/2K$. The  energy at the maximum is a barrier preventing phase slips:
\be 
E_b =\pi AM_\perp^2\ln {1\over 2Kr_c}. 
    \ee{}
The barrier vanishes if $K$ becomes of the order of the inverse vortex core radius. This conclusion is applicable also to antiferromagnets. 

The core radius $r_c$ is of order of the coherence length determined from the spin wave spectrum as was indicated earlier. However, different modes have different coherence lengths, and it is necessary to understand which kind of a vortex corresponds to which spin wave mode. In vortex cores spins form skyrmions.
Variation of magnetization vectors  in the skyrmion vortex core as a function of the distance $r$ from the vortex axis is shown schematically in Fig.~\ref{SC}. In the ferromagnet [Fig.~\ref{SC}(a)]  there is only one spin wave mode, and the radius of the core is determined by the coherence length $\xi_0$ for this mode [see \eq{xi0}]. In the antiferromagnet there are two spin wave modes and, correspondingly, there are two types of vortices. The skyrmion core connected to the Goldstone mode is illustrated in Fig.~\ref{SC}(b). Only the pair of the angle variables ($\theta$,$\varphi$) vary inside the core, while $\Theta=\Phi=0$. Figure~\ref{SC}(c) illustrates the skyrmion core connected to the gapped mode. The magnetization vectors $\bm M_1$ and $\bm M_2$ rotate around the axis normal to the magnetic field, and there are nonzero $\Theta$ and $\Phi$. The core radius is determined by the coherence length $\xi$ given by \eq{xi}.

\section{Long-distance superfluid spin transport } \label{SpTr} 

From the time when the concept of spin superfluidity was suggested\cite{ES-78b}, it was debated about whether the superfluid spin current is a ``real'' transport current. As a response to this concern, in Ref.~\onlinecite{ES-78b}  a {\em Gedanken} (at that time) experiment for demonstration of reality of superfluid spin transport was proposed  (see also more recent Refs.~\onlinecite{Adv,Tserk,Halp}). 

\begin{figure}[t]
\includegraphics[width=.7\textwidth]{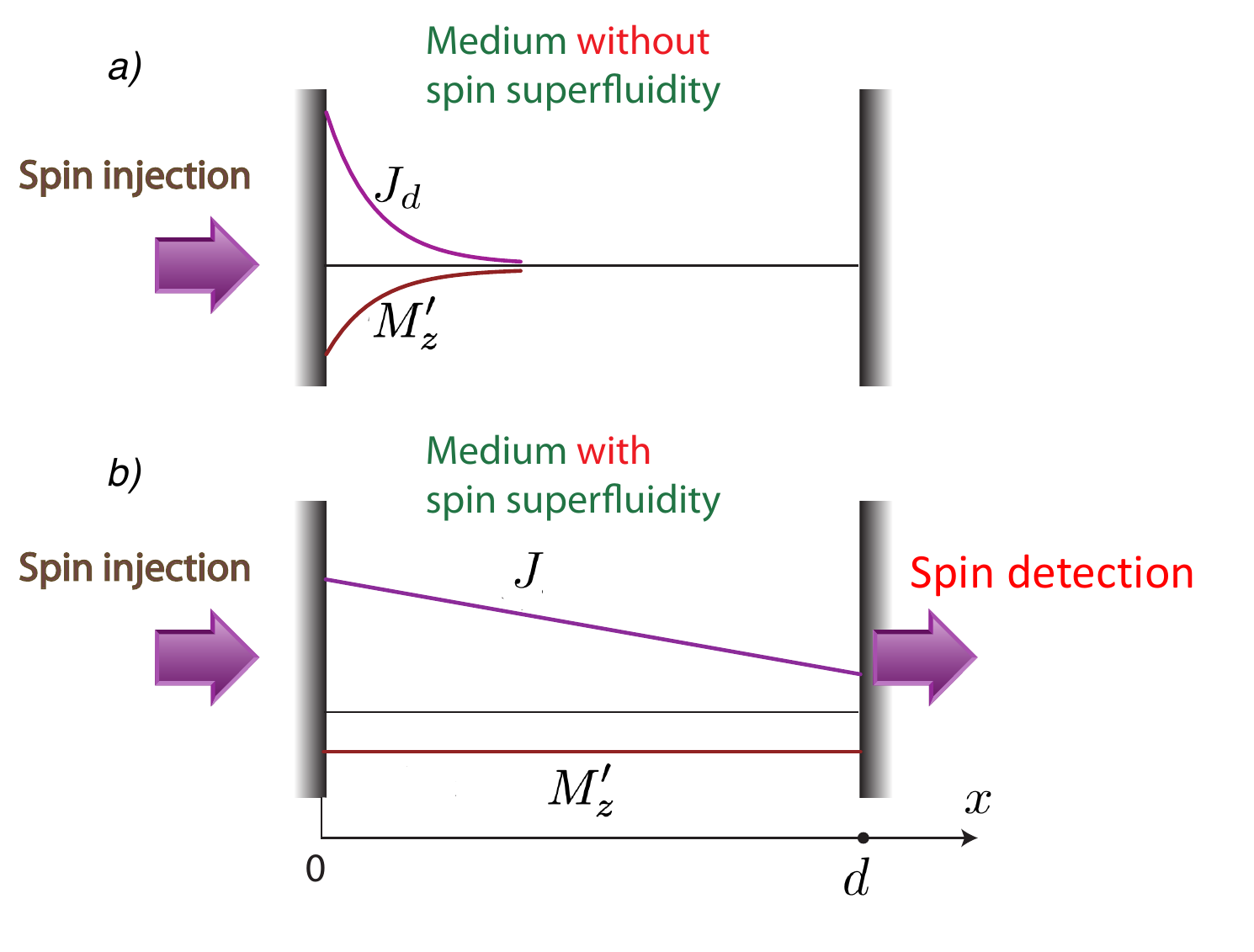}
\caption[]{ Long distance spin transport. (a) Spin injection to a spin-nonsuperfluid medium. (b) Spin injection to a  spin-superfluid medium.  }
\label{f1}
\end{figure}

The spin is injected to one side of a magnetically ordered layer of thickness $d$ and spin accumulation is checked at another side (Fig.~\ref{f1}). 
For the analysis of spin transport in this set up we must modified the continuity equation (\ref{Em}) for the ferromagnet adding two dissipation terms:
     \be
{1\over \gamma} {dM_z \over dt}=-\bm \nabla \cdot \bm J-\bm \nabla \cdot \bm J_d- {M'_z \over \gamma T_1},
      \ee{EmB}
Here $M'_z=M_z-\chi H$ is the non-equilibrium magnetization  and  the superfluid spin current $\bm J$ is given by \eq{cur}. The first dissipation term is the spin diffusion current 
    \be
 \bm  J_d =- {D\over \gamma}\bm\nabla M_z.
       \ee{Jd}
Spin diffusion does not violates the spin conservation law. The second dissipation term is connected with the longitudinal spin relaxation. It is characterized by the Bloch time $T_1$ and does violate the spin conservation law.

 In the absence of spin superfluidity ($\bm J=0$) \eq{Ep} for the spin phase is not relevant, and \eq{EmB} describes pure spin diffusion [Fig.~\ref{f1}(a)].  Its solution, with the boundary condition that the spin current $J_0$ is injected at  the interface $x=0$, is
\be
J_d=J_0e^{-x/L_d}, ~~M'_z=\gamma J_0\sqrt{T_1\over D}  e^{-x/L_d},
   \ee{}
where 
\be
L_d=\sqrt{DT_1} 
   \ee{}
is the spin-diffusion length. Thus, the effect of spin injection exponentially decays at the scale of the spin-diffusion length,  and the density of  spin accumulated at the other side of the layer decreases exponentially  with growing distance $d$.

However, if spin superfluidity is possible, the spin precession equation (\ref{Ep}) becomes relevant. According to this equation, in a stationary state the magnetization $ M'_z$ cannot vary in space [Fig.~\ref{f1}(b)]  since according to \eq{Ep} the gradient $\bm\nabla M'_z$ is accompanied by the linear in time growth of the gradient $\bm\nabla \varphi$. 
The right-hand side of \eq{Ep} is an analog of the chemical potential, and the requirement of constant in space magnetization $M_z$  is similar to the requirement of constant in space chemical potential in superfluids, or the electrochemical potential in superconductors.  As a consequence of this requirement, spin diffusion current is impossible in the bulk since  it is simply ``short-circuited'' by the superfluid spin current. The bulk spin diffusion current can appear only in AC processes.

If the spin superfluidity is possible, the spin current can reach the spin detector at the plane $x=d$  opposite to the border where spin is injected.  As a boundary condition at $x=d$, one can use a phenomenological relation connecting the spin current with the magnetization:
$J(d) = M'_z(d) v_d$, where $v_d$ is a phenomenological constant. This boundary condition was confirmed by the microscopic theory of \citet{Tserk}. Together with the boundary condition $J(0) = J_0$ at $x=0$ this yields the solution of Eqs.~(\ref{Ep}) and (\ref{EmB}):
 \be
M'_z= { T_1 \over d+v_dT_1}\gamma J_0,~~J(x) = J_0 \left(1-{x\over d+v_d  T_1} \right).
     \ee{}
Thus, the spin accumulated at large distance $d$ from the spin injector slowly decreases with $d$ as  $1/(d+C)$ [Fig.~\ref{f1}(b)], in contrast to the exponential decay $\propto e^{-d/L_d}$ in the spin diffusion transport [Fig.~\ref{f1}(a)]. The constant $C$ is determined by the boundary condition at $x=d$.

\section{Experimental detection of spin superfluidity} \label{DC}

\begin{figure}[t]
\includegraphics[width=.5\textwidth]{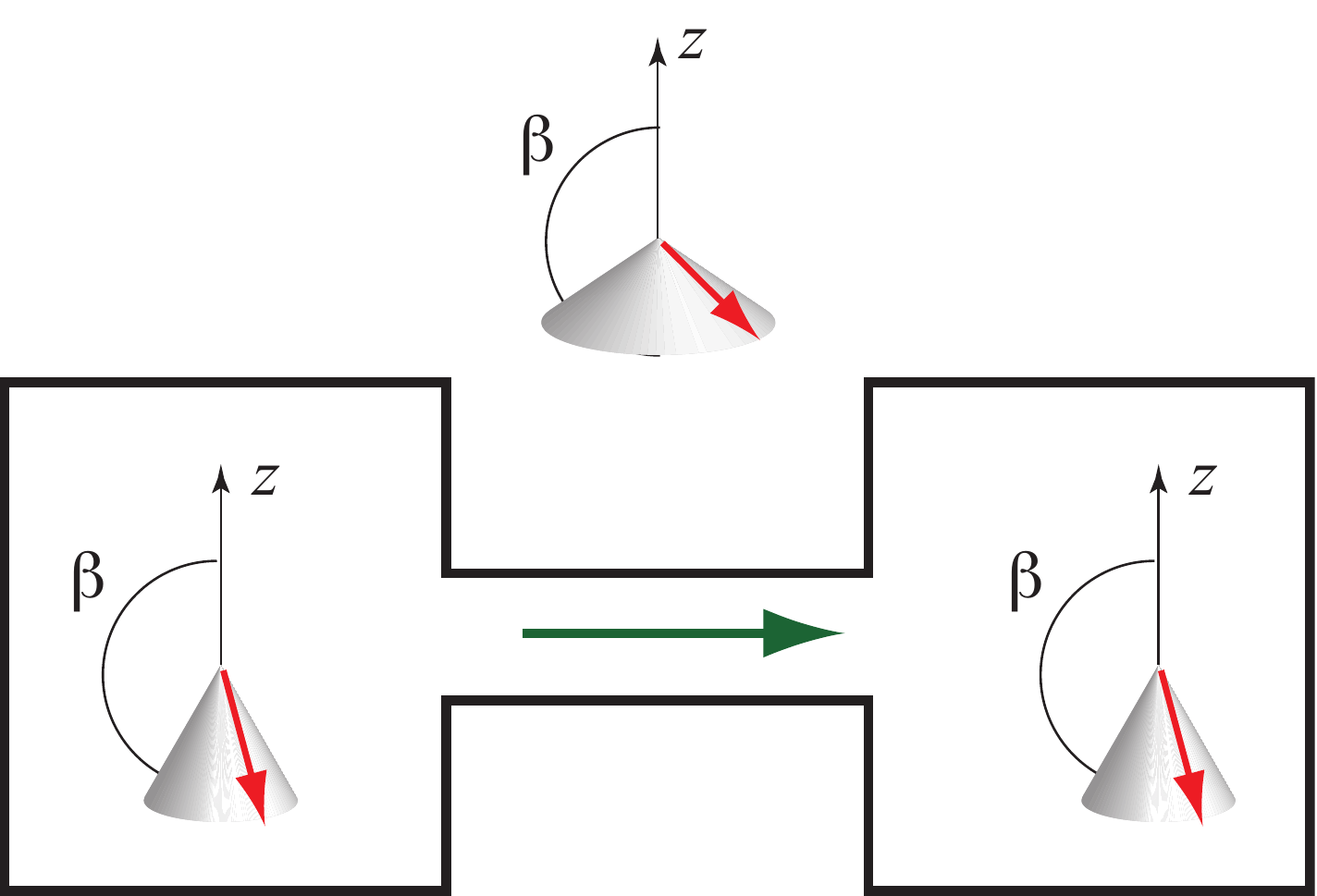}
\caption[]{Spin transport through a channel connecting  two cells filled by the $B$ phase of superfluid $^3$He. The horizontal  arrow shows the direction of the spin current in the channel.\cite{flux}  }
\label{fig4.1}
\end{figure}

A smoking gun of the possibility of spin supercurrents in the $B$-phase of superfluid $^3$He was  an experiment with a spin current through a long channel connecting two cells filled by the superfluid $^3$He.\cite{flux}  The quasi-equilibrium state of the coherent spin precession (later rebranded as magnon BEC\cite{BunV})  was supported by spin pumping. The magnetic fields applied  to the two cells were slightly different, and therefore, the spins in the two cells precessed  with different frequencies. A small difference in the frequencies  leads to a linear growth of
difference of the precession phases in the cells and a phase gradient in the channel. When the gradient reached the critical value, $2\pi$ phase slips were detected in the experiment. 
The sharp $2\pi$  phase slip  was reliable evidence of non-trivial spin supercurrents at phase gradients restricted by finite critical values. 

 It was important evidence that persistent spin currents are possible.  However, real long-distance transportation of spin by these currents was not demonstrated.  Moreover, it is impossible to do in the non-equilibrium magnon BEC, which was realized in the $B$ phase of $^3$He superfluid\cite{BunV} and in yttrium-iron-garnet magnetic films.\cite{Dem6} The non-equilibrium magnon BEC  requires  pumping of spin in the whole bulk for its existence. In the geometry of the aforementioned spin transport experiment this would  mean that spin is permanently pumped not only by a distant injector but also all the way up the place where its accumulation is probed. Thus, the spin detector measures  not only spin coming from the distant injector but also  spin pumped close to the detector. Therefore, the experiment does not prove the existence of long-distance spin superfluid transport. 
 
\begin{figure}[t]
\includegraphics[width=.7\textwidth]{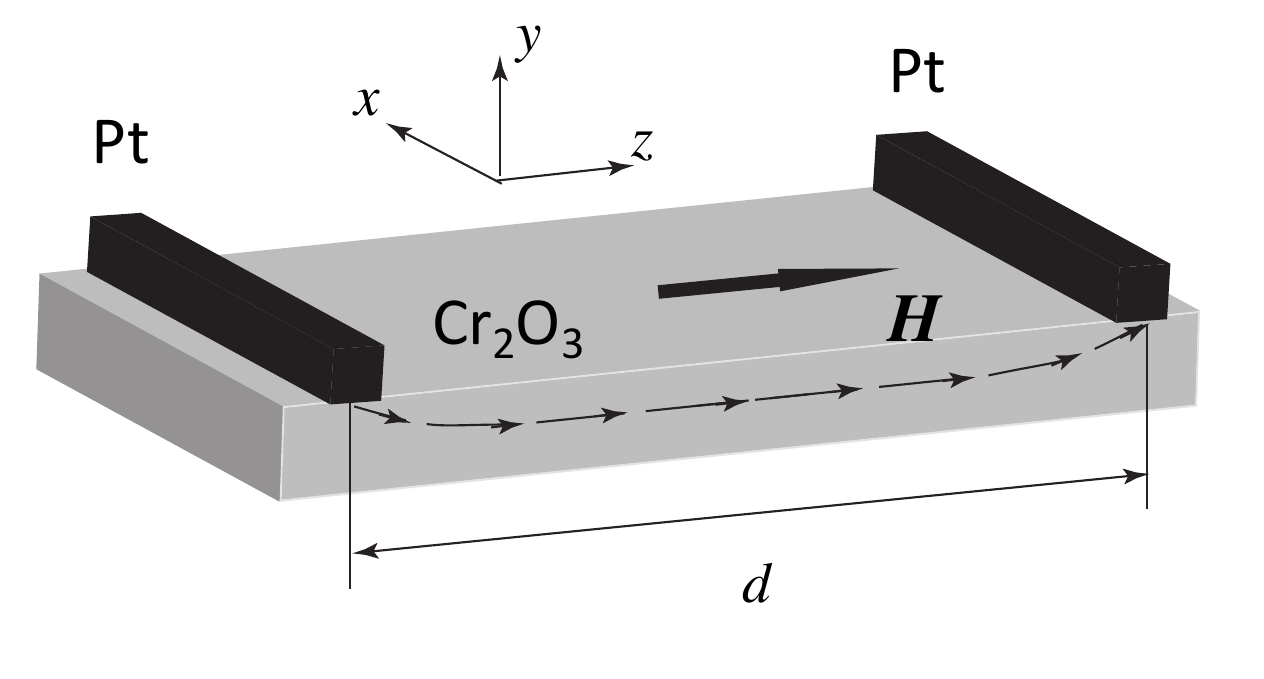}
\caption[]{ Long distance spin transport in the geometry of the experiment by \citet{WeiH}.
Spin is  injected from the left Pt wire and flows along the Cr$_2$O$_3$ film  to the right Pt wire, which serves as a detector. The arrowed dashed line shows a spin-current streamline. }
\label{fy}
\end{figure}

The experiment suggested for detection of long-distance superfluid spin transport\cite{ES-78b}  was recently done by \citet{WeiH} in antiferromagnetic Cr$_2$O$_3$. In the experiment of \citet{WeiH} the spin is created  in the Pt injector by heating (the Seebeck effect) on one side of the Cr$_2$O$_3$ film and spin accumulation is probed on another side of the film by the Pt detector via the inverse spin Hall effect (Fig.~\ref{fy}). In agreement with theoretical prediction, they observed spin accumulation inversely proportional to the distance from the interface where spin was injected into  Cr$_2$O$_3$.  

In Fig.~\ref{f1} the spin flows along the axis $x$, while the spin and the magnetic field are directed along the axis $z$. In the geometry of the experiment of \citet{WeiH} the spin flows along the spin axis $z$ parallel to the magnetic field. This geometry is shown in Fig.~\ref{fy}. The difference between two geometries is not essential if spin-orbit coupling is ignored. In our theoretical analysis we chose the geometry with different directions of the spin current and the spin in order to stress the possibility of the independent choice of axes in the spin and the configurational spaces. 
  
 There were other reports on experimental detection of spin superfluidity in magnetically ordered solids. \citet{spinY} declared detection of spin superfluidity at high temperatures in a decaying magnon condensate in a YIG film.   In their experiment the phase gradient emerged because  of  spin precession difference produced by a temperature gradient. However, the estimate made in Ref.~\cite{Son17} showed that the total phase difference across the magnon cloud in the experiment  did not exceed  about 1/3 of the full $2\pi$ rotation. Thus,  \citet{spinY} could detect only  microscopic  spin currents emerging in any spin wave.  As explained above, ``superfluidity'' connected with such currents was well proved by numerous half-a-century old experiments on spin waves at {\em all} temperatures and does not need new experimental confirmations.  

Observation of the long-distance  superfluid spin transport was also reported by \citet{Lau18} in a graphene quantum Hall antiferromagnet.  However, the discussion of this report requires an extensive theoretical analysis of the $\nu=0$ quantum Hall state of graphene, which goes beyond the scope of the present review. A reader can find this analysis in Ref.~\onlinecite{SupGraph}.

\section{Discussion and conclusions}\label{disc}

The paper addressed the basics of the spin superfluidity concept: topology, Landau criterion, and phase slips. Metastable (persistent) superfluid current states are possible if the order parameter space (vacuum manifold) has the topology of a circumference on a plane like in conventional superfluids. In ferromagnets it is the circumference on the spherical surface in the space of spontaneous magnetizations $\bm M$. In antiferromagnets it is the circumference on the unit sphere in the space of the unit N\'eel vector $\bm L/L$, where  $\bm L$ is the staggered magnetization. The topology necessary for spin superfluidity requires the uniaxial easy-plane anisotropy in ferromagnets, while in antiferromagnets this anisotropy is provided by the Zeeman energy, which confines the N\'eel vector in the plane normal to the magnetic field.

The Landau criterion was checked  for the spectrum of elementary excitations, which are spin waves in our case. In ferromagnets there is only one Goldstone spin wave mode. In bipartite antiferromagnets there are two modes: the Goldstone mode in which spins perform rotational oscillations around the symmetry axis and the gapped mode with rotational oscillations around the axis normal to the symmetry axis. At weak magnetic fields the Landau instability starts not in the Goldstone mode, but in the gapped mode.  In contrast to superfluid mass currents in conventional superfluids, metastable spin superfluid currents are restricted not only by the Landau criterion from above but also from below. The restriction from below is related to the absence of the strict conservation law for spin. 

The Landau instability with respect to elementary excitations is a precursor for the instability with respect to phase slips.  The latter instability starts  when the spin phase gradient reaches the value of the inverse vortex core radius. This value is on the same order of magnitude  as the Landau critical gradient.  Vortices participating in phase slips have skyrmion cores, which map on the upper or lower part of the spherical surface in the space of spontaneous magnetizations in ferromagnets, or in the space of the unit N\'eel vectors  in antiferromagnets. 

It is worthwhile to note that in reality it is not easy to reach  the critical gradients discussed in the present paper experimentally. The decay of superfluid spin currents is possible also at subcritical spin phase gradients since the barriers for phase slips can be overcome by thermal activation or macroscopic quantum tunneling.  This makes the very definition of the real critical gradient rather ambiguous  and dependent on duration of observation of persistent currents.   Calculation of real critical gradients requires a detailed dynamical analysis of  processes of thermal activation or macroscopic quantum tunneling through phase slip barriers, which is beyond the scope of the present paper. One can find examples of such analysis for conventional superfluids with mass supercurrents in Ref.~\onlinecite{EBS}.
  
Although evidence of the existence of metastable superfluid spin currents in the $B$ phase of superfluid $^3$He  were reported long ago\cite{flux}  the first experiment demonstrating  the long-distance transport of spin by these currents in the solid antiferromagnet was done only recently.\cite{WeiH} This is not the end but the beginning of the experimental verification of the long-distance superfluid spin transport in magnetically ordered solids. 
In the experiment of \citet{WeiH}  spin injection required  heating of the Pt injector,  and the spin current to the detector is inevitably accompanied by a heat flow. \citet{Lebrun} argued that probably \citet{WeiH} detected a signal not from spin coming from the injector but from spin generated by the Seebeck effect at the interface between the heated antiferromagnet and the Pt detector. Such effect has already been observed for antiferromagnet Cr$_2$O$_3$.\cite{Seki} If true,  \citet{WeiH} observed not long-distance spin transport but long-distance heat transport. However, it is not supported by the fact that  Yuan {\em et al.} observed a threshold for  superfluid spin transport at low intensity of injection, when according to the theory (see Sec.~\ref{Phase}) the absence  of the strict spin conservation law becomes important.  With all that said, the heat-transport interpretation cannot be ruled out and deserves further investigation. According to this interpretation, one can see the signal observed by \citet{WeiH} at the detector even  if the Pt injector is replaced by a heater, which produces the same heat but no spin. An experimental check of this prediction would confirm or reject the heat-transport interpretation.   

The present paper focused on spin superfluidity  in  magnetically ordered solids. Recently investigations of spin superfluidity were extended to spin-1 BEC, where spin and mass superfluidity coexist and interplay.\cite{LamSpin,Duine,Sp1,Sp1af}
This interplay leads to a number of new nontrivial features of the phenomenon of superfluidity. The both types of superfluidity are restricted by the Landau criterion for the softer  
collective modes, which are the spin wave modes. As a result, the presence of spin superfluidity diminishes the possibility of the conventional mass superfluidity. Another  consequence of the coexistence of spin and mass superfluidity is phase slips with bicirculation vortices characterized by two topological charges (winding numbers).\cite{Sp1af}


%

\end{document}